\documentclass{aastex631}
\usepackage{multirow}
\newcommand{\tess}{{\it TESS}}

\begin{document}
\title{The Changing Lightcurve of the Double-Mode RR Lyrae Variable Star V338 Boo}
\correspondingauthor{Kenneth Carrell}
\email{kenneth.carrell@angelo.edu}

\author[0000-0002-6307-992X]{Kenneth Carrell}
\affiliation{Physics \& Geosciences,
  Angelo State University\\
  2601 W.\ Avenue N
  San Angelo, TX  76909}

\author{Ronald Wilhelm}
\affiliation{Department of Physics \& Astronomy,
  University of Kentucky\\
  177 Chem.-Phys.\ Building,
  505 Rose Street,
  Lexington, KY  40506}

\author{Faith Olsen}
\affiliation{Physics \& Geosciences,
  Angelo State University\\
  2601 W.\ Avenue N
  San Angelo, TX  76909}

\author{Andrew Tom}
\affiliation{Physics \& Geosciences,
  Angelo State University\\
  2601 W.\ Avenue N
  San Angelo, TX  76909}

\author{Garath Vetters}
\affiliation{Physics \& Geosciences,
  Angelo State University\\
  2601 W.\ Avenue N
  San Angelo, TX  76909}

\author{Anna McElhannon}
\affiliation{Department of Physics \& Astronomy,
  University of Kentucky\\
  177 Chem.-Phys.\ Building,
  505 Rose Street,
  Lexington, KY  40506}

\accepted{to \apjl\ July 7, 2021}

\begin{abstract}
  We present an analysis of the lightcurve extracted from Transiting
  Exoplanet Survey Satellite Full Frame Images of the double-mode RR
  Lyrae V338 Boo. We find that the fundamental mode pulsation is
  changing in amplitude across the 54 days of observations. The first
  overtone mode pulsation also changes, but on a much smaller
  scale. Harmonics and combinations of the primary pulsation modes
  also exhibit unusual behavior. Possible connections with other
  changes in RR Lyrae pulsations are discussed, but a full
  understanding of the cause of the changes seen in V338 Boo should
  shed light on some of the most difficult and unanswered questions in
  stellar pulsation theory, and astrophysics more generally.
\end{abstract}

\section{Introduction}\label{sec:intro}
V338 Boo was first reported as a double-mode RR Lyrae star (RRd) by
\citet{oaster06} using the Northern Sky Variability Survey
\citep[NSVS,][]{nsvs} data. Although the period ratio of the two modes of
pulsation fell within the normal range ($P_1/P_0$ = 0.743), the ratio
of the amplitudes of the two pulsations was found to be strange - the
fundamental mode amplitude was twice as large as the first overtone
mode. Follow-up observations were published in 2010 \citep{jaavso} and
2012 \citep{jaavso2} which showed that the amplitude ratio was
changing over the course of several years.

Given the ID 5222076 in the NSVS and the name V338 Boo by the General
Catalog of Variable Stars \citep[GCVS,][]{gcvs}, the Transiting
Exoplanet Survey Satellite \citep[\tess,][]{tess} observed this
star with a \tess\ input catalog (TIC) ID of 282941520 in Sectors 23
(camera 3) and 24 (camera 2), which covered the dates March 18 - April
16, 2020 and April 16 - May 13, 2020, respectively. These observations
provide two major advantages over ground-based observations obtained
previously. The most obvious is the increased precision of the
photometric measurements. Arguably more important, however, is the
continuous monitoring of the star for almost two months.

In this Letter we present an analysis of \tess\ Full Frame Images
(FFIs) of V338 Boo showing the changing amplitude ratio is
still ongoing, and we provide a more detailed description of its
behavior.

\section{Analysis}\label{sec:analysis}
Initial data analysis was performed with the python package 
\verb|A TESS Archive RR Lyrae Classifier|
\citep[\texttt{ATARRI},][]{atarri}. This graphical user
interface (GUI) uses the \verb|search_tesscut| functionality
\citep{tesscut} provided by the \verb|lightkurve| python package
\citep{lightkurve} to download cutouts of the \tess\ FFIs and present
the data and preliminary analysis for quick visual inspection.
\texttt{ATARRI} is listed on the Astrophysics Source
  Code Library (ASCL) website and provides the user with the ability
  to verify the quality of the cutout image and eliminate bad data
  points from the presented lightcurve. It also has the Fourier
  analysis of the lightcurve visible, as well as a folded
  lightcurve. This GUI is meant to quickly and easily present
  information to the user in order to classify the RR Lyrae star into
  a type, and to look for behavior that would require a more detailed
  follow-up analysis. More information about the \texttt{ATARRI}
  package is available in the ASCL and in the GitHub repository
  hosting the code.
The initial inspection of V338 Boo indicated 
a detailed follow-up was warranted.

The full analysis of V338 Boo used the same python packages as
\verb|ATARRI| to download FFI data. A threshold mask was used to
determine the correct aperture and background from the cutouts and to
extract lightcurves from the data. We convert the background
subtracted flux to a magnitude using the value provided in the
TIC of Tmag = 12.5954.

\begin{figure}[ht]
  \plotone{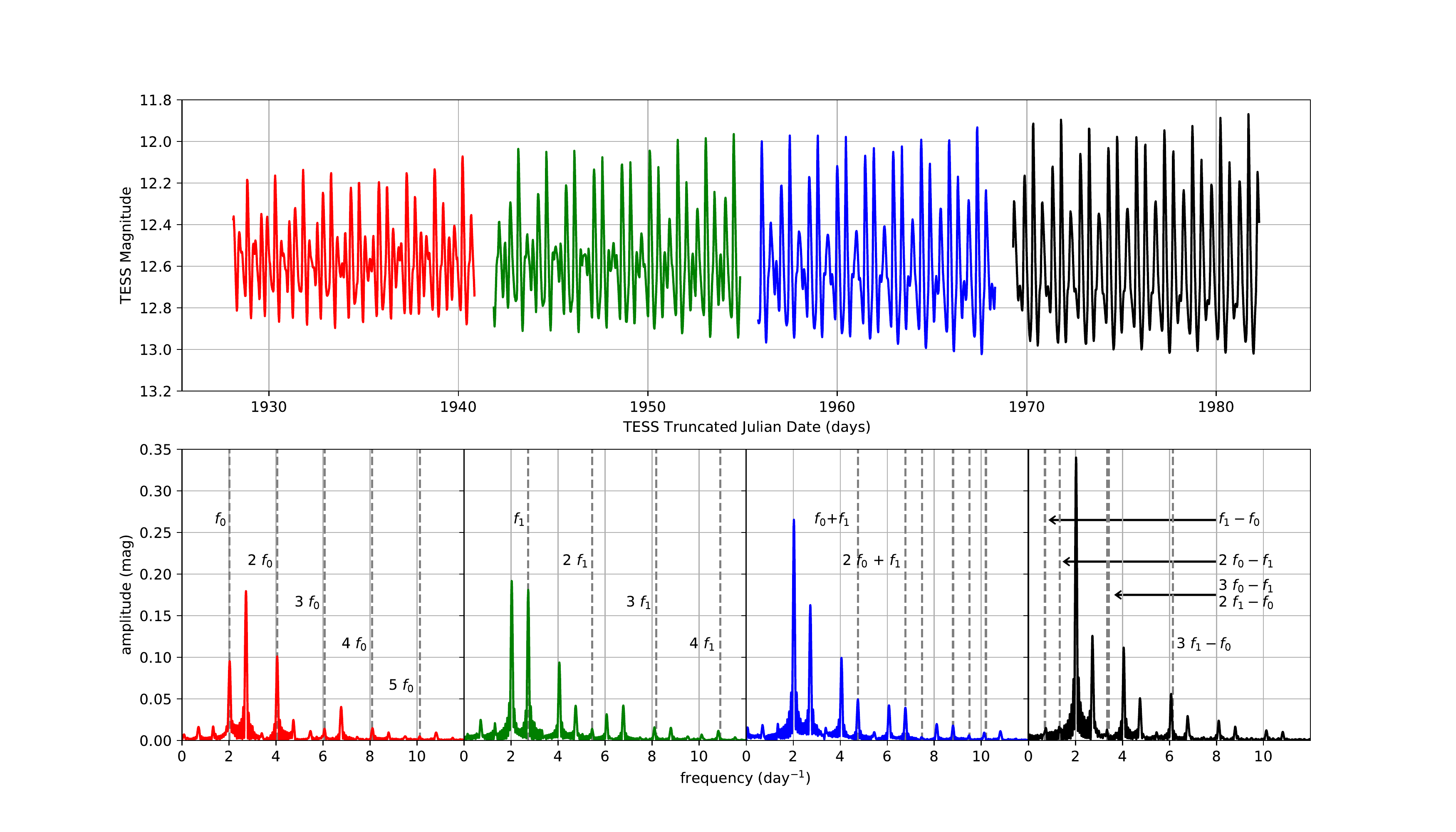}
  \caption{Top: The full lightcurve of V338 Boo. Bottom: Frequency
    analysis of the color-coded segments from the upper plot. Dashed
    lines and text label the locations of various peaks, with $f_0$
    and $f_1$ corresponding to the fundamental and first overtone
    frequencies, respectively.}
  \label{fig:segments}
\end{figure}
The top panel of Figure \ref{fig:segments} shows the full lightcurve
of V338 Boo from \tess\ Sectors 23 and 24. There is a
definite change in both the pulsation amplitude and shape of the lightcurve 
across the 54 days it was observed. 
In the bottom panels of Figure \ref{fig:segments} we
show a frequency analysis of the data using \verb|Period04|
\citep{period04}. The data 
has been segmented and is color-coded between the upper and lower
plots. The changing pulsation cycles in the upper plot can be more
clearly seen in the frequency analysis of the segments. In particular,
the fundamental mode amplitude peak at frequency 2.02464 d$^{-1}$ more
than doubles between the first and last segments, and changes from a
secondary peak to the dominant one compared to the first overtone mode
peak at 2.72454 d$^{-1}$.

\begin{figure}[ht]
  \plotone{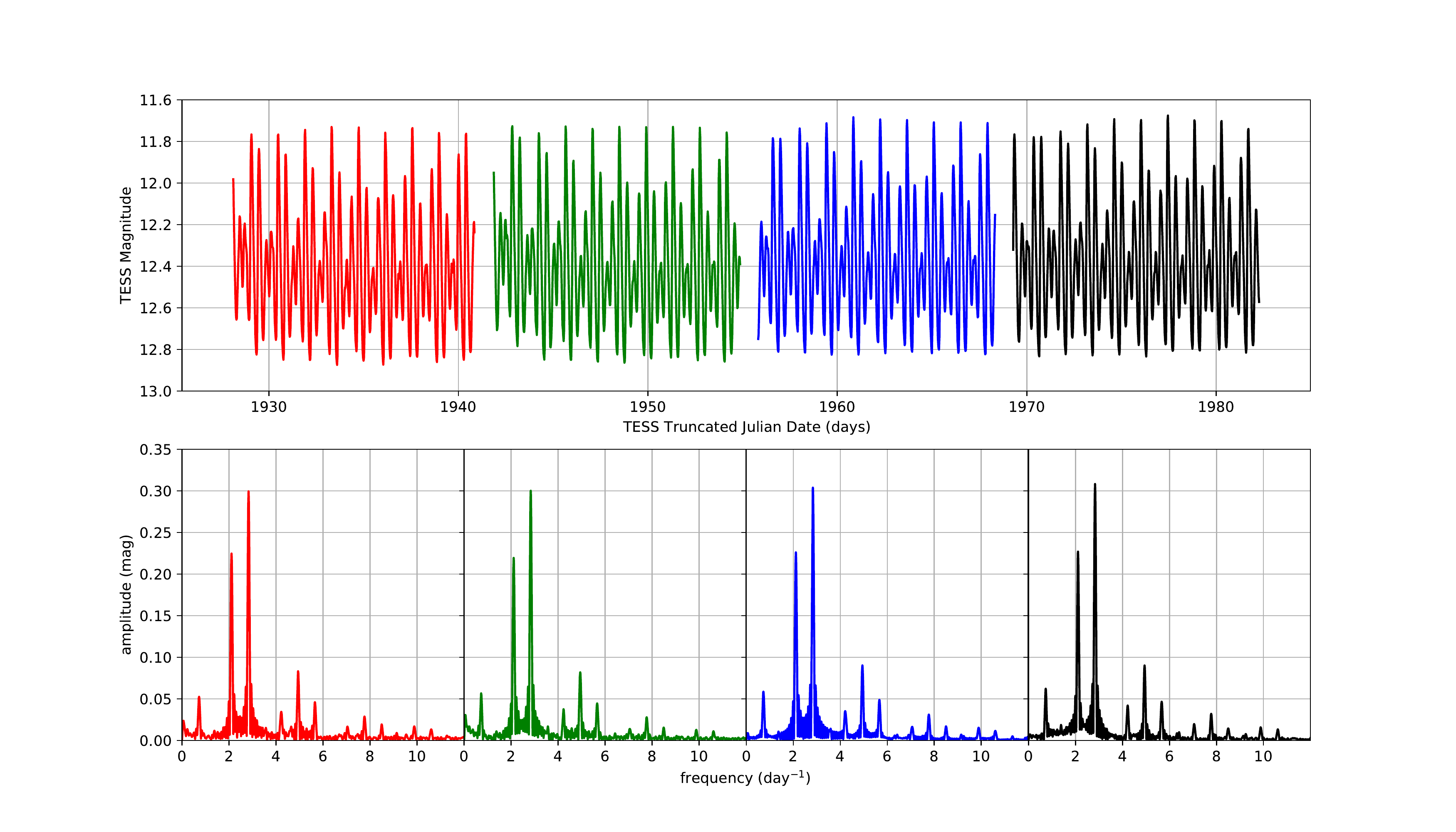}
  \caption{The same plots as Figure \ref{fig:segments} for the normal
    RRd variable NN Boo.}
  \label{fig:segComp}
\end{figure}
As a comparison, we performed the same analysis on the RRd star NN
Boo, which has a similar \tess\ magnitude (12.387) and was observed in
the same sectors as V338 Boo (but in different cameras - 2 and 1
respectively for Sectors 23 and 24). Figure \ref{fig:segComp} shows
the analysis for NN Boo. The behavior of NN Boo is typical of a
normal RRd variable - the first overtone pulsation is dominant
in the frequency analysis, and the roughly 3:4 ratio of the
frequencies of the pulsation
modes is evident in the pulsation pattern of the stable
lightcurve. Most 
importantly for this analysis, the peak amplitudes for both the
fundamental and first overtone modes are constant.

\begin{figure}[ht]
  \plotone{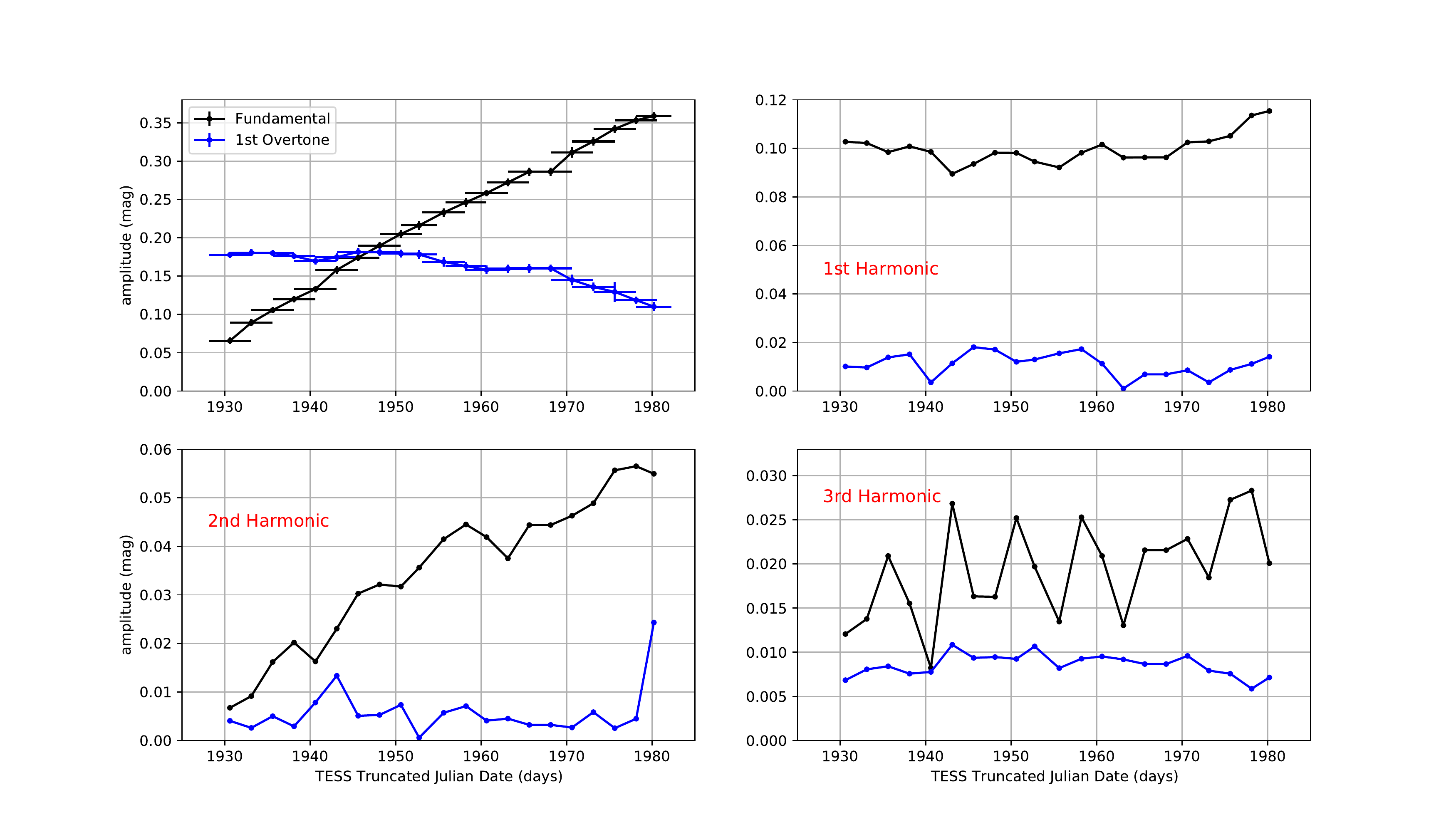}
  \caption{The amplitude of the fundamental mode (black) and first
    overtone mode (blue) for the primary peak (top left), first
    harmonic (top right), second harmonic (bottom left), and third
    harmonic (bottom right) for V338 Boo.}
  \label{fig:amp}
\end{figure}

To examine the change in the lightcurve of V338 Boo in more detail, we
performed a frequency analysis of samples of five day increments,
shifting by 2.5 days. In Figure \ref{fig:amp} we show how the
amplitudes of the primary frequencies (top left) and next three
harmonics change over time.
The change of the fundamental mode from a secondary peak to the
primary peak is clearly evident. The fundamental mode peak changes
from a value of 0.07 mag in the first five days of observations to a value
of 0.36 mag in the last five days. This factor of five increase is much
larger than the decrease experienced by the first overtone mode peak,
which starts with a value of 0.18 mag and ends with a value of
0.11 mag.

The harmonics of the fundamental mode display peculiar trends. The
first harmonic does not change in amplitude during these
observations. The second harmonic, however, changes by almost a factor
of ten - it goes from 0.006 to 0.055 mag.

Additionally, the peak amplitude of the first harmonic of the
fundamental mode is {\it larger} than the amplitude of the
fundamental mode peak itself for an extended period in the beginning
of the observations. This can also be seen in the frequency analysis
of the first segment of data in Figure \ref{fig:segments}.

An analysis of the frequency of the primary peaks in the five day
samples shows that there is small variations but no obvious trends in
the data similar to those seen in Figure \ref{fig:amp}. Furthermore,
our frequencies of $f_0$ = 2.02464 $\pm$ 0.00008 d$^{-1}$ and $f_1$ =
2.72454 $\pm$ 0.00018 d$^{-1}$ agree well with the results of
\citet{oaster06}  - our $P_0$ = 0.4939 d and $P_1$ = 0.3670 d compared
to their $P_0$ = 0.4940 d and $P_1$ = 0.3668 d.

\begin{figure}[ht]
  \plotone{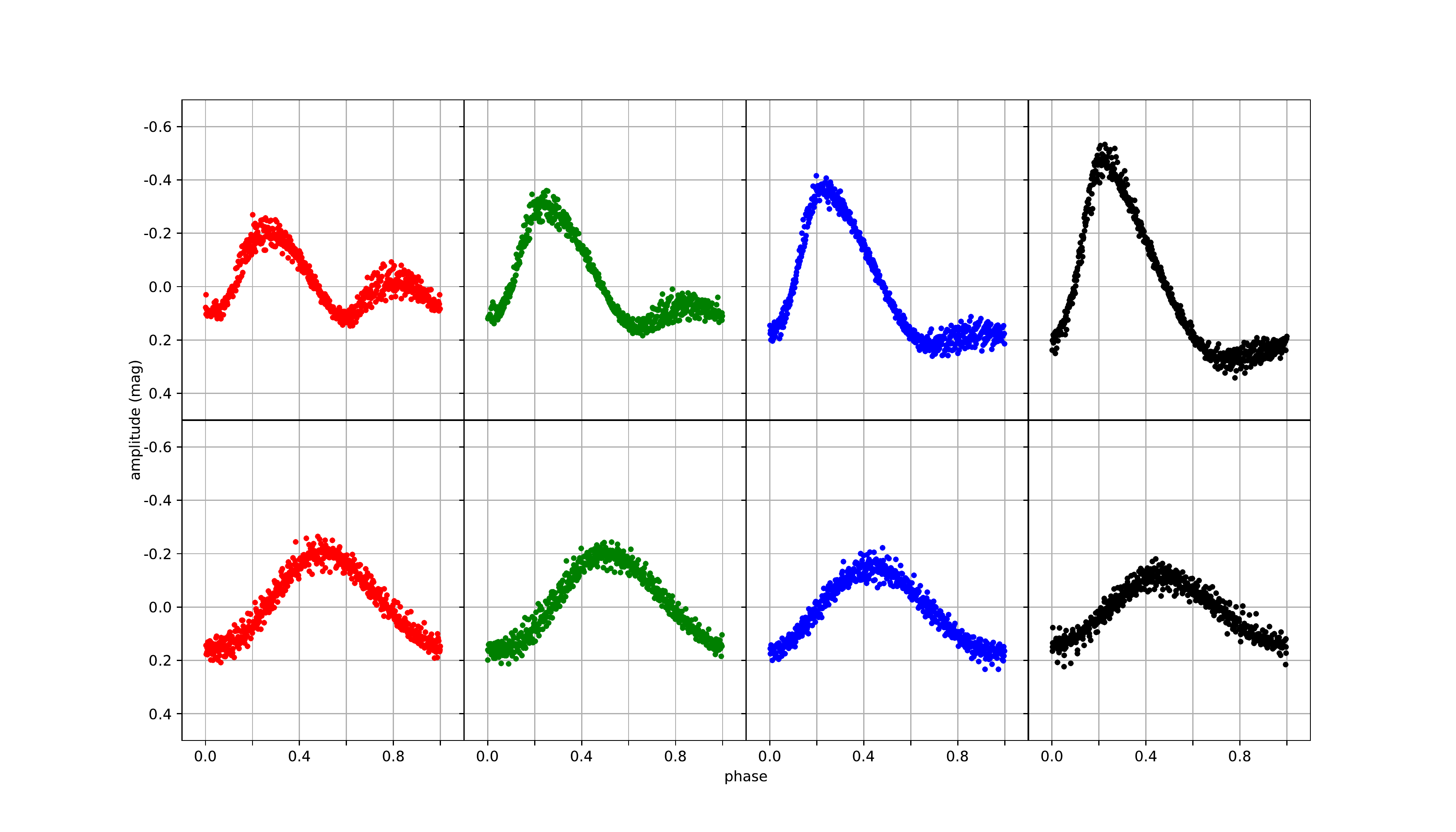}
  \caption{The folded lightcurve of each segment of V338 Boo after
    subtracting out the first overtone mode plus combined peaks to show
    the fundamental mode only (top row), and subtracting out the
    fundamental mode plus combined peaks to show the first overtone
    mode only (bottom row). Colors and order match those from Figure
    \ref{fig:segments}.}
  \label{fig:separated}
\end{figure}
As a further analysis, we examine the separated, folded lightcurves
of each segment of the V338 Boo \tess\ data. For each segment, we
fit each pulsation mode and the next four harmonics for each, and
combinations of the two modes for those with frequencies less than
approximately 8 d$^{-1}$.  In the top row of Figure
\ref{fig:separated} is the fundamental mode folded lightcurve
obtained by subtracting out the first overtone mode, its next four
harmonics, and the combined frequencies. The phase has been shifted
by the same constant offset for each segment for clarity. The bottom
row shows the first overtone mode obtained after subtracting the
fundamental mode, its next four harmonics, and the combined
frequencies. The phase for these have been shifted by the same
constant offset for each segment for clarity as well, although with
a different offset than the fundamental mode. The fundamental mode
lightcurve grows in amplitude going from left to right, which is
what is seen in the Fourier analysis presented above. Also, the
first segment of data (top left plot of Figure 
\ref{fig:separated}) has a pronounced bump at a phase of $\sim$
0.8. As the amplitude grows, this bump decreases. The presence of
the bump is what causes the first harmonic peak of the fundamental
mode to be large in the first segment, and as the bump disappears,
the growing amplitude of the fundamental mode pulsation balances out
this peak in the Fourier analysis. The folded lightcurves for the first
overtone mode have very slight changes in comparison, again matching
what is seen above. There is a small decrease in the amplitude, and
there appears to be a small phase shift of the peak.

The changes in the lightcurve of V338 Boo in some ways resemble
changes seen by \citet{cha14} in the lightcurve of the RRab variable
S Arae. They saw the amplitude of a bump in the lightcurve and the
overall shape of the lightcurve change as a function of the Blazhko
phase for the star. The changes seen here are similar, and are also
in the fundamental mode pulsation, which could mean we are seeing
the same phenomenon in the RRd star V338 Boo. Follow-up observations
and further analysis are required to confirm this.

Because of the large pixel size of \tess, contamination from
other sources can be a major concern. The contamination ratio
calculated for V338 Boo in the TIC is 0.1109609 and there are three
objects within one arcminute of the star. Two of the three are much
fainter than V338 Boo: one at 14.6 arcseconds with a \tess\
magnitude of 17.68, and one at 57.6 arcseconds
with a \tess\ 
magnitude of 17.74. The star 24.0 arcseconds
away has a \tess\ 
magnitude of 14.08, which is about one and a
half magnitudes 
fainter than V338 Boo. The next closest star with a \tess\ magnitude
brighter than 15 is over two arcminutes away. Contamination from
non-variable sources will contribute extra flux and affect amplitude
changes. However, they will not change the frequency analysis or
overall properties of its changes. If there were a nearby variable
source contributing significantly to the lightcurve of V338 Boo, we
would see additional peaks in the Fourier analysis. However, the
fundamental and first overtone pulsation peaks we find agree with
previous results, and we find no other significant peaks in the
Fourier analysis that are not associated with these two modes, their
harmonics, and their combinations, which gives us confidence that
contamination from nearby sources does not significantly affect our
results.

\section{Discussion}\label{sec:disc}
\subsection{Is this new?}\label{ssec:new}
An obvious question arises in light of this new data: Is the behavior
new or was V338 Boo changing this way in previous data as well?

As an attempt to answer this question, we randomly sampled the \tess\
data in a manner that would be similar to ground-based
observations. Since observing from the ground only allows for roughly
half day windows, we randomly selected 20 days from the full data set
and took the first half day of observations for each day. A frequency
analysis was performed on these random 20 half-day samples and the
amplitude ratio of the fundamental mode to first overtone mode peaks
are given in Table \ref{tab:randos}. This table also provides the same
ratio measured by previous studies. We note that our 20 ``night''
sample is more than the follow-up observations of \citet{oaster06} (16
nights), \citet{jaavso} (13 and 4 nights), and \citet{jaavso2} (10 and
12 nights), and that the window of observations of \tess\ (54 days) is
shorter than those taken previously (between 66 and 108 days). With
the integration time of \tess\ FFIs (30 mins) being 
much longer than these previous results, however, we increased the
number of days in our random samples to give more points for the
frequency analysis. Table \ref{tab:randos} compares the amplitude
ratios of the previous results with those from our random samples. The
full range of amplitude ratios is reproduced in our random
samples, but we generally find lower ratios than those reported
previously. The lightcurves of RR Lyrae can exhibit 
differences in different passbands, notably in the infrared
\citep{jur18}, but the \tess\ passband encompasses the I band (and
extends both bluer and redder), which makes comparison between those
amplitude ratios and the \tess\ ones in Table \ref{tab:randos} more
meaningful.
We therefore suggest that V338
Boo has been exhibiting transient, fast (tens
or hundreds of days) changes in the fundamental mode amplitude since
detailed observations of it began in 
2005 at the very least. It is also likely that we are not
seeing the full range of change in our sample of two \tess\
sectors. The higher amplitude ratios of previous results suggest that
either we are not seeing the maximum amplitude ratio in these
observations, and/or that the change of the
fundamental mode peak amplitude spends more time at the higher values
than the lower.
\begin{table}[ht]
  \centering
  \begin{tabular}{| l r | c |}
    \hline
    \multicolumn{2}{|c|}{{\bf previous results}} &
                                                 {\bf random samples
                                                   from \tess} \\
    \hline\hline
    \multicolumn{2}{|c|}{{\bf V band values}} & {\bf Tmag values}\\
    \citet{oaster06} & 1.93 & 1.21 \\
    \multirow{2}{*}{\citet{jaavso}} & 1.76 $\pm$ 0.03 (2008) & 1.24 \\
                                                 & 1.48 $\pm$ 0.01
                                                   (2009) & 1.25 \\
    \multirow{2}{*}{\citet{jaavso2}} & 1.40 $\pm$ 0.02 (2010) & 1.28 \\
                                                 & 1.82 $\pm$ 0.02
                                                   (2011) & 1.36 \\ \cline{1-2}
    \multicolumn{2}{|c|}{{\bf I band values}} & 1.38 \\
    \multirow{2}{*}{\citet{jaavso}} & 1.56 $\pm$ 0.06 (2008) & 1.43 \\
                                                 & 1.52 $\pm$ 0.03
                                                   (2009) & 1.45 \\
    \multirow{2}{*}{\citet{jaavso2}} & 1.38 $\pm$ 0.03 (2010) & 1.69 \\
                                                 & 1.81 $\pm$ 0.03
                                                   (2011) & 1.88 \\
    \hline
  \end{tabular}
  \caption{A comparison of the amplitude ratios, A$_0$/A$_1$ (where 0
    and 1 represent the fundamental and first overtone modes, 
    respectively) found in previous works with
    those from random samples taken from the \tess\ data created to
    simulate ground-based observations.}
  \label{tab:randos}
\end{table}

\subsection{Possible interpretations}\label{ssec:interp}
We have shown that what was previously believed to be slow changes in
the lightcurve of V338 Boo over the course of years is actually
changes on the timescale of several tens of days. Obviously,
interpretations of physical mechanisms behind this behavior must now
be revisited.

Although there are few examples, RR Lyrae have been found to switch
their pulsation modes in the past. The variable V79 in M3 was seen to
change modes in 1992 \citep{v79} and then change back to its original
mode in 2007 \citep{v79back}. The Optical Gravitational Lensing
Experiment (OGLE) has found four examples of RR Lyrae changing modes
\citep{ogle1,ogle2,ogle3} and the Catalina Sky Survey (CSS) found six
\citep{css}. The commonality found between all of these mode-switching
RR Lyrae is they all appear to change from a double-mode pulsation
(RRd) to a fundamental mode only pulsation (RRab) or vice versa. It is
tempting to attribute the behavior we see to 
mode-switching, or an intermediate state of it. However, V338 Boo has
always been observed to have both pulsation modes present, and in
previous cases of mode-switching the time of fundamental mode only
pulsations showed no signs of first overtone pulsations. Our window of
observations from \tess\ ends with the fundamental mode still
increasing in amplitude, so we cannot rule out the possibility that
there was a full mode-switch to an RRab type after these
observations. Given the fact that it appears the change in fundamental
mode peak amplitude has been ongoing since 2005, we find it unlikely
that this set of observations occurred just before a complete change in
that behavior.
If V338 Boo is experiencing changes in its pulsation due to
evolutionary effects \citep[e.g.][]{buc02}, the timescale
for that change is much longer than both what was seen in previous
mode-switching stars and in the thermal timescale.  Follow-up
observations are necessary to determine the physical properties of
this star in order to compare with the narrow range of parameter
space seen in stable double-mode pulsations from hydrodynamic
simulations \citep{sza04} and what that means for
its evolutionary stage.

Some double-mode RR Lyrae have been labeled as ``anomalous'' by some
authors for a variety of reasons. \citet{2Omode} find 22 examples of RRd
stars in the Magellanic Clouds with properties different than those of
``regular'' RRd type. Specifically, they find period ratios of the
first overtone to fundamental mode below what is typically found,
fundamental mode peak amplitudes larger than the first overtone mode,
and long-term changes of the amplitudes of the pulsation modes.
While V338 Boo has a period
ratio in the ``normal'' range for RRd, the fundamental mode peak is
larger than the first overtone in previous results and in most of the
\tess\ observation window. Additionally,
analysis of the \tess\ lightcurve shows both a changing
fundamental and first overtone mode pulsation
amplitude. 
It is important to note that \citet{2Omode} 
find changes in amplitude by looking for sidebands to the
radial mode frequencies, which are indicative of the Blazhko effect
\citep{blazhko} since
the sidebands are formed from the combination of the pulsation and
Blazhko peaks. Of their sample of 22 anomalous RRd, 18 have radial
mode frequencies with sidebands, and 11 of those were found to have
modulations in the fundamental mode exclusively. Even though the
\tess\ observation window for V338 Boo is too short to make a clear
determination of sidebands, as previously discussed, the peak
amplitude of the fundamental mode changes significantly during the
\tess\ observations.
If V338 Boo were a single-mode pulsator (RRab or
RRc) it would undoubtedly be classified as a Blazhko
type. Additionally, the behavior reported here is in some 
ways similar to the modulation effects seen by \citet{smo15}. However,
as in \citet{2Omode}, their sample has lower period ratios than
typical RRd variables and is from a specific population (the Galactic bulge).

A few facts about this dataset are obvious. First, although the
primary peak of the fundamental mode pulsation is dramatically
changing and the second harmonic of the peak also changes, the first
harmonic of the fundamental mode changes very little. Second, 
the combination peak $f_0 +
f_1$ increases in a similar fashion to the primary fundamental mode
peak. Also, the $2 f_0 + f_1$ peak
at a frequency of $\sim$6.77 d$^{-1}$ is quite large - it is the
fourth highest peak in the first segment of data in Figure
\ref{fig:segments}. Strangely, however, this combination peak does not
increase in amplitude with the fundamental mode change, and it
actually {\it decreases} in the last segment of data when the
fundamental mode is the largest. There is also a peak at a frequency
of $\sim$3.4 d$^{-1}$ that corresponds to the combination $2f_1 -
f_0$, although it has a low S/N. In \citet{2Omode}, it was claimed
that the anomalous behavior of RRd stars in the Magellanic Clouds
could be due to a resonance with the second overtone radial pulsation
mode. In this scenario, the resonance $\Delta f = 2 f_{1O} -
f_F - f_{2O} = 0$. This gives $f_{2O} = 2 f_{1O} - f_F$, which
corresponds to the peak we see at $\sim$3.4 d$^{-1}$, and the first
harmonic of this peak ($2 f_{2O}$) would be approximately at our peak
at $\sim$6.77 d$^{-1}$. Therefore, the constant peak at $\sim$6.77
d$^{-1}$ could potentially be caused by an increasing contribution of
the fundamental mode coupled with a decreasing contribution of the
second overtone mode.

\section{Conclusion}\label{sec:con}
We have presented the lightcurve of V338 Boo from \tess\ FFI data. A
Fourier analysis of this RRd star shows several peculiar behaviors.
\begin{itemize}
  \item The fundamental mode pulsation peak changes dramatically over
    the course of these observations - starting as a secondary peak and
    becoming the dominant peak. To our knowledge, this is the first
    time a change of this magnitude and on this timescale has
    been reported, and the first 
    time there has been a change reported in the dominant
    pulsation mode of an RRd with a normal period ratio.
  \item The change seen here in the fundamental mode can explain
    changes in the ratio of the amplitudes of the fundamental to first
    overtone modes seen over several years in previous studies.
  \item Despite the large change in the fundamental mode peak, its
    first harmonic remains relatively constant during the same period.
  \item Some combination peaks of the fundamental and first overtone
    modes also display changing amplitudes, but there are notably some
    that do not, including the peak at $2 f_0 + f_1$.
\end{itemize}

The relative brightness of V338 Boo makes it a perfect star for 
follow-up observations necessary to more fully describe the
behavior seen here and to explore possible explanations. 
Photometry is needed to ensure
that a full mode-switch from RRd to RRab did not occur after these
observations. \tess\ will reobserve this star in Cycle 4, which can be
used to verify and extend these results or determine that a full
mode-switch has occurred. Spectroscopy will aid by showing how the
outer layers of this star are physically moving, and if there is any
anomalous behavior in that motion. Caution must be taken in any
follow-up observations, however,
because as was seen with previous results, the period of
pulsations and the change seen in the fundamental mode peak
amplitude will affect the results.

An explanation for the strange behavior exhibited by V338 Boo could
potentially shed light on some of the longest and most difficult
unanswered questions in all of astrophysics. In particular, these
changes could be related to the Blazhko effect, which would provide a
new window into how this effect behaves. On the other hand, if V338
Boo is a mode-switching variable, its apparent
transient behavior 
could set physical constraints on the manner and timescale of the
mode-switching phenomenon. Whichever the case, V338 Boo is an unusual
and exciting star for future observations.

\begin{acknowledgments}
  We wish to thank the anonymous reviewer for suggests that made this
  manuscript better.
  
  Funding was provided to KC through a grant from the
  Angelo State University Faculty Research Enhancement
  Program.

  This paper includes data collected by the TESS mission.
  Funding for the TESS mission is provided by the NASA's
  Science Mission Directorate.
\end{acknowledgments}

\bibliography{paper}{}
\bibliographystyle{aasjournal}

\end{document}